\documentclass[twocolumn,showpacs,preprintnumbers,amsmath,amssymb]{revtex4}
\usepackage{graphics}
\usepackage{epsfig}
\usepackage{color}
\usepackage{subfigure}

\begin{document}

\title{Out-of-equilibrium statistical ensemble inequivalence}
\author{G. De Ninno$^{1,2}$, D. Fanelli$^{3}$}
\affiliation{1. Sincrotrone Trieste, S.S. 14 km 163.5, Basovizza (Ts), Italy \\
2. Physics Department, Nova Gorica University, Nova Gorica, Slovenia \\
3. Dipartimento di Energetica "S. Stecco" and INFN, University of Florence, Via S. Marta 3, 50139 Florence, Italy
}

\date{\today}

\begin{abstract}
We consider a paradigmatic model describing the one-dimensional motion of $N$ rotators coupled through 
a mean-field interaction, and subject to the perturbation of an external magnetic field. The latter is shown to significantly alter the system behaviour, driving the emergence of ensemble inequivalence in the out-of-equilibrium phase, as signalled by a negative (microcanonical) magnetic susceptibility. The thermodynamic of the system is analytically discussed, building on a maximum entropy scheme justified from first principles. Simulations confirm the adequacy of the theoretical picture. Ensemble inequivalence is shown to rely on a peculiar phenomenon, different from the one observed in previous works. As a result, the existence of a convex intruder in the micro-canonical energy is found to be a necessary but not sufficient condition for inequivalence to be (macroscopically) observed. Negative temperature states are also found to occur. These intriguing phenomena reflect the non-Boltzmanian nature of the scrutinized problem and, as such, bear traits of universality that embraces equilibrium as well out-of-equilibrium regimes.
\end{abstract}

\maketitle

Classical statistical mechanics is most commonly dealing with large systems, in which the interaction range among components is much smaller than the system size. In such ``short-range"  systems, energy is normally additive and statitistical ensembles are equivalent \cite{book}. The situation may be radically different when the interaction potential decays so slowly that the force experienced by any system element is dominated by the interaction with far-away components. In these ``long-range" interacting systems (LRI) energy is not additive. Well-known examples of non-additive LRI systems are for instance found in cosmology (self-gravitating systems) and plasma physics applications, where Coulomb interaction are at play \cite{bookLRI, reviewRuffo}. The lack of additivity, together with the possible non-convexity of the space of accessible macroscopic thermodynamic parameters (break of ergodicity), may be at the origin of a large gallery of peculiar thermodynamic behaviours: the specific heat can be negative in the microcanonical ensemble  \cite{reviewRuffo}, and temperature jumps may appear at microcanonical first-order phase transitions. When this occurs, experiments realized on isolated systems give different result from similar experiments performed on systems in contact with a thermal bath. As a consequence, the canonical and microcanonical statistical ensembles of long-range interacting systems may be non-equivalent.

Ensemble inequivalence has been theoretically observed in several models of LRI systems, at thermodynamical equilibrium \cite{bookLRI}. In all cases, for isolated systems such an observation has been related to the a convex shape of the microcanonical entropy \cite{gross}. Indeed, if the entropy $s(e)$ is twice differentiable and non-concave for some values of the specific energy $e$, the microcanonical specific heat is negative. Thus, since the canonical (thermodynamic) specific heat is a positive-defined quantity, ensembles are non-equivalent. In open systems, $s$ may also depend on other extensive thermodynamic variables, apart from energy. In this case, (equilibrium) ensemble inequivalence has been detected also if $s$ is non-convex along $e$, provided it is convex along one of the other variables \cite{CampaRuffoTouchette}.
   
Systems with LRI also display a quite peculiar dynamics, which is characterized by a slow relaxation towards thermodynamical equilibrium, and, even more remarkably, by the convergence to out-of-equilibrium quasi-stationary states (QSS) \cite{Yamaguchi, AntoniazziPRL1,fel}. It has been shown that QSS's can be related to the stable steady states of the Vlasov equation, which describes the system in the limit $N \rightarrow \infty$  \cite{reviewRuffo}. The idea was inspired by the seminal work of Lynden-Bell  \cite{LyndenBell68}, developed in the context of stellar dynamics, and later applied to vortex dynamics (see, e.g.,  \cite{chavanis}). As it will be discussed more in detail in the following, Lynden-Bell's approach is based on the definition of a locally-averaged (``coarse-grained") distribution function, which then translates into an entropy functional, as follows from standard 
statistical mechanics prescriptions. By maximizing such an entropy, while imposing the constraints of the dynamics, returns a closed analytical expression for the  
single particle distribution of the system in its QSS regime. The microcanonical Lynden-Bell entropy enables one to introduce  
an out-of-equilibrium free-energy, defined via the usual Legendre-Frenchel transform \cite{reviewRuffo,PRLstaniscia}. This makes it possible 
to estimate the relevant collective variables, as calculated both in the microcanonical and canonical out-of-equilibrium statistical ensembles,  to eventually challenge 
the equivalence of the two formulations. Following this procedure, it was recently demonstrated \cite{PRLstaniscia} that (out-of-equilibrium) ensemble equivalence may still hold when negative kinetic specific heat is measured in both statistical ensembles. While QSS's represent out-of-equilibrium states of the $N$-body dynamics, they could be equally interpreted as equilibrium configuration of the corresponding continuous description: in this respect, the conclusions of our analysis will apply to both equilibrium and non equilibrium dynamics, provided the latter bears distinctive 
non-Boltzmannian traits.  

In this letter, we extend the concept of statistical inequivalence to out-of equilibrium LRI systems. As it will be shown, the inequivalence does not materialize as an immediate byproduct of a ``convex intruder" in the microcanonical entropy. Indeed, the entropy is always a convex function of one of the thermodynamic variables, over the whole domain of its definition, including the region where experiments realized in the two ensembles would provide the same results (ensemble equivalence). This observation, that we here substantiate via numerical and analytical means, suggests that the presence of a ``convex intruder" in the entropy is a necessary but not sufficient condition for the inequivalence of statistical ensembles to be observed in a macroscopically tangible form, e.g. as a difference in sign of the system's specific heat and/or susceptibility. This phenomenon should be therefore categorized as a new type of inequivalence, 
distinct from the one so far reported in the literature. Further, the inequivalence is here shown to emerge as an effect of an externally imposed forcing, that significantly impacts the associated, unperturbed dynamics for which equivalence is shown to hold \cite{PRLstaniscia}. Finally, the results here derived can be also cast in the framework of an equilibrium non-Boltzmanian picture, as in the spirit of 
\cite{PRLstaniscia}. 

As a paradigmatic example for our investigation, we consider a model describing the one-dimensional motion of $N$ rotators coupled through 
a mean-field cosine interaction and subject to the perturbation of an external magnetic-like field. The model is mathematically defined by the following Hamiltonian:

\begin{equation}
\label{eq:ham}
H_0 = \frac{1}{2} \sum_{j=1}^N p_j^2 + \frac{1}{2 N} \sum_{i,j=1}^N
\left[1 -  \cos(\theta_j-\theta_i) \right] -\frac{h}{N}\sum_{j=1}^N\cos(\theta_j). 
\end{equation}

Here $\theta_j$ represents the orientation of the $j$-th rotor and $p_j$ is its conjugate momentum. The parameter $h$ is a scalar quantity, which measures the strength of the magnetic field. For $h=0$, the previous Hamiltonian reduces to the celebrated Hamiltonian Mean Field (HMF) model \cite{antoni-95}, which has been widely studied in the past as a prototype of LRI system. To monitor the evolution of the
system, it is customary to introduce the magnetization, a macroscopic order
parameter defined as $M=|{\mathbf M}|=|\sum {\mathbf m_i}| /N$, where
${\mathbf m_i}=(\cos \theta_i,\sin \theta_i)$ stands for the microscopic
magnetization vector. 

As previously reported \cite{antoni-95}, for $h=0$, after an initial transient, the system gets trapped into QSS's. 
Similarly, QSS's are also encountered in the generalized setting with $h \neq 0$.

In the $N \to \infty$ limit, i.e. when the system is indefinitely stuck in the QSS's, the discrete dynamics generated by the Hamiltonian (\ref{eq:ham}) can be described in terms of a continuum Vlasov equation. The QSS's are then interpreted as the stable (attractive) steady states of the underlying Vlasov equation. Analytical progress is possible by invoking the aforementioned Lynden--Bell violent relaxation theory \cite{LyndenBell68}, that we shall tackle with reference to a simplified choice of the initial condition. 
Assume that the initial single particle distribution takes only two distinct values, namely $f_0$ and zero. Angles and velocities populate a bound domain in phase space, therein displaying a uniform probability distribution. This working ansatz corresponds to dealing with the so-called ``water-bag" distribution. Vlasov time evolution can modify the shape of the boundary of the ``water-bag", while conserving the area inside it. Thus, the distribution remains two-level ($0,f_0$) as time
progresses. By performing a local average of $f$ inside a given mesoscopic box, one gets a coarse-grained distribution  $\bar{f}$, which eventually converges to an asymptotic equilibrium profile. The latter solution is explicitly calculated by maximising an entropy functional, associated to $\bar{f}$, through a direct combinatorial 
calculation that moves from a rigorous statistical mechanics setting \cite{LyndenBell68}. 
In the two-level scenario, the mixing entropy $s$ can be cast in the form: 
\begin{equation}
\label{entropy_shape}
s(\bar{f})=-\int \!\!{\mathrm d}p{\mathrm d}\theta \,
\left[\frac{\bar{f}}{f_0} \ln \frac{\bar{f}}{f_0}
+\left(1-\frac{\bar{f}}{f_0}\right)\ln
\left(1-\frac{\bar{f}}{f_0}\right)\right].
\end{equation}

If one operates in the setting of Hamiltonian (\ref{eq:ham}), both the specific energy $
\epsilon(\bar{f}) = \int \int (p^2/{2}) f(\theta,p,t) {\mathrm d}\theta {\mathrm d}p - ({M_x^2+M_y^2 - 1})/{2}- h M_x$
and the momentum $P(\bar{f}) = \int \int p f(\theta,p,t) {\mathrm d}\theta {\mathrm d}p$
are conserved by the dynamics. In addition, the normalization of the distribution $\bar{f}$ has to be 
accounted for, which in turn corresponds to dealing with constant mass. 
Requiring the entropy to be stationary, while imposing the conservation of the above quantities 
($\epsilon(\bar{f}) = e$ and $P(\bar{f}) = \sigma$), defines a variational problem that admits as a solution the following fermionic distribution:

\begin{multline}
  \label{eq:barf}
  \bar{f}_{\text{QSS}}(\theta,p)= \\
  \frac{f_0}{ 1+e^{\displaystyle\beta f_0 (p^2/2 -{\mathbf M} [\bar{f}_{\text{QSS}}] \cdot {\mathbf m} -h\cos\theta)+\alpha}},
\end{multline} 

where the label ${\text{QSS}}$  recalls that the recovered Vlasov equilibrium distribution is meant to describe the out-of-equilibrium (QSS) $N$-body regime. Here $\alpha$ and $\beta$ play the role of Lagrange multipliers associated, respectively, to mass and energy conservation; ${\mathbf M}=(M_x, M_y)$, where $M_x[\bar{f}]=\int \bar{f} \cos(\theta) d\theta d p$, $M_y[\bar{f}]=\int \bar{f} \sin(\theta) d\theta d p$ stand for the two components of magnetization in the $N\rightarrow \infty$ limit; finally, ${\mathbf m}=(\cos \theta,\sin \theta)$. 
Once the system energy $e$ and $f_0$ are fixed, one can determine the corresponding values of $M_x$, $M_y$, $\alpha$ and $\beta$ in the expression (\ref{eq:barf}), by self-consistently imposing the constraints condition. 
The obtained system of equations is then solved using a Newton-Raphson method.

\begin{figure}
\centering
\includegraphics[width=5cm]{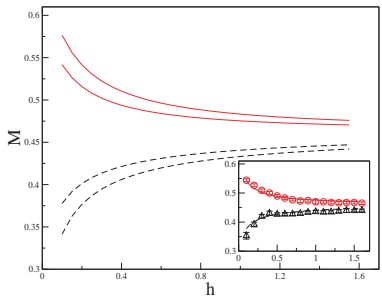}
\caption{Magnetization $M$ in the QSS, calculated using the Lynden-Bell approach, plotted as a function of $h$, for different choices of $e$ (continuous and segmented lines). From top to bottom: $e=0.51,0.53,0.62,0.64$. In the inset, the results of direct $N$-body simulations (symbols) are compared to the Lynden-Bell prediction, for $e=0.53$ and $e=0.62$.  
In the simulations, $N=10000$. Each data point refers to an average over 20 independent realizations. The error bar is calculated as the variance of the 20 independent estimates.}
\label{fig1}
\end{figure}

We are in particular interested in monitoring the response of the system to the external solicitation here encoded in the parameter $h$. Fixing the energy, while changing $h$ returns a sequence of distinct equilibrium values of the magnetization, as calculated via the above maximum entropy procedure. It is worth noting that the stationary entropy depends parametrically on both the energy and (quasi-stationary) magnetization, the latter being independent thermodynamic quantities. 
The magnetization values are displayed in fig.\ref{fig1} as a function of $h$, and for different energy choices. It is worth emphasising that we here operate at constant $f_0$. Surprisingly, below a threshold in energy the system shows negative susceptibility $\chi = \partial M / \partial h$, the magnetization decreasing with the strength of the externally imposed field. Above the critical energy value, the magnetization grows instead with $h$, which implies dealing with a positive susceptibility.

The presence of a region with negative susceptibility in the microcanonical system (\ref{eq:ham}), points to out-of-equilibrium ensemble inequivalence. Imagine in fact to introduce the system free energy, by performing the Legendre-Frenchel transform of entropy (\ref{entropy_shape}) \cite{PRLstaniscia}. In the so-defined canonical ensemble, the sign of $\chi$ is related to the concavity of the free energy. The latter being (by definition) always concave, the canonical susceptibility is positive defined. In other words, $\chi$ can be negative in the microcanonical ensemble, as it does
in a specific parameters region,  while it is always positive definite quantity in the canonical framework. The
theoretical evidence for $\chi<0$ represents hence an indirect (macroscopically accessible) signature of ensemble inequivalence. 

To confirm the adequacy of the proposed theoretical picture we performed a campaign of direct N-body simulations, sampling the QSS, after the initial transient has died out. The magnetization is computed as a function of $h$, averaging over many independent realizations. The recorded data are 
plotted in the inset of fig.\ref{fig1}, showing a quite satisfactory agreement with the analytical calculation. 

In the remaining part of the paper, we shall elaborate on a thermodynamical description of the phenomenon here detected, by exploring the topological properties of the entropy functional (\ref{entropy_shape}) versus energy and magnetization.

\begin{figure}
\centering
\includegraphics[width=5cm]{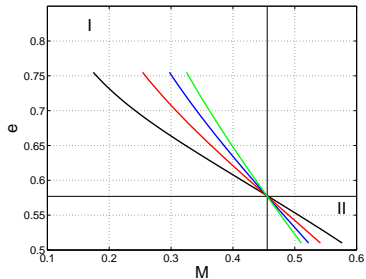}
\caption{Projection of $s(e,M)$ in the $(M,e)$ plan: each line is an iso-$h$, with $h=0.1,0.2,0.3,0.4$, from bottom to top in region I (vice versa in region II).
The lines intersect in a unique point, $(M_c,e_c) \simeq (0.46,0.575)$. Here, $f_0=0.12$. Regions I and II as defined in the text are visually delimited by two solid, orthogonally crossing, lines.}
\label{fig2}
\end{figure}

For a magnetic system, the the first principle of thermodynamics can be cast in the form \footnote{This form of the principle is 
straightforwardly adapted from  \cite{Amit}, eq. (2.1.10): the change $h \rightarrow -h$ is operated as follows a consistency argument 
to make contact with our definition of the energy (see Hamiltonian (\ref{eq:ham})) where 
the external perturbation enters with a negative sign.}

\begin{equation}\label{eqTdS}
T d s = d e + h dM 
\end{equation}
 
where $T$ is the thermodynamic temperature and $e$ is the total energy of the system, {\it including the field contribution} . Recalling 
that $s$ is a function of the independent variables $(e,M)$ one readily obtains the following expression for $h$:

\begin{equation}\label{eqh}
h = \frac{\left( \frac{\partial s}{\partial M} \right)_{e}}{\left( \frac{\partial s}{\partial e} \right)_{M}}. 
\end{equation}

where use has been made of the relation $T=\left( \partial s/\partial e \right)_M^{-1}$. From the expression for $h$, one can immediately cast the susceptibility in the micro-canonical statistical ensemble as:  

\begin{equation}\label{eqchi}
\chi = \left( \frac{\partial h}{\partial M} \right)^{-1} = 
\left( \frac{\partial s}{\partial e} \right)^2 \left( \frac{\partial s}{\partial e}  \frac{\partial^2 s}{\partial M^2} - 
\frac{\partial^2 s}{\partial  M \partial e}  \frac{\partial s}{\partial M} \right)^{-1}, 
\end{equation}

where for the sake of clarity we dropped the label that identifies the variable kept constant upon differentiation.
The sign of $\chi$, and thus the associated (macroscopically accessible) out-of-equilibrium inequivalence features, are ultimately  
controlled by the (first, second and crossed) partial derivatives of the entropy $s$ with respect to $M$ and $e$. Using the Lynden-Bell approach, the three dimensional profile 
of $s$ can be analytically reconstructed as a function of the energy $e$ and (self-consistently determined) magnetization $M$. In fig. \ref{fig2} we report a 2D projection of the entropy iso-$h$ lines in the $(M,e)$ plan, for $f_0=0.12$: each line refer to a specific value of $h$, which is frozen to the selected value, while visiting the entropy surface. The lines cross in a well defined point of the plan that we identify as 
$(M_c,e_c) \simeq (0.46,0.575)$ \footnote{Notice that the values of $M_c$ and $e_c$ depends on the specific $f_0$ employed. The same qualitative behaviour is however found for different choices of $f_0$.}. Two different domains can be singled out, which, as will become transparent 
later on, appear to be disconnected: a larger region, termed I in the following, delimited by the conditions $e > e_c$ and $M < M_c$. A smaller region II, in which $e < e_c$ and $M > M_c$ \footnote{The energy is limited from below, the minimum value that it can eventually attain depending on $f_0$, see \cite{PRLstaniscia}.}. 

\begin{figure}
\centering
\includegraphics[width=8cm]{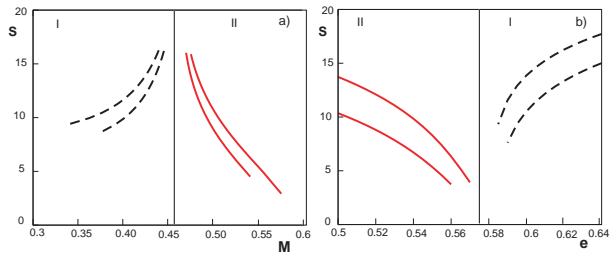}
\caption{Fig. a): Entropy $s$ is plotted versus (QSS) magnetization $M$, for different choices of the energy: $e = 0.51, 0.53, 0.62, 0.64$, from left to right. Fig. b): $s$ is plotted versus $e$, for different choices of the (QSS) magnetization $M$: $M=0.49, 0.51, 0.45, 0.43$, from left to right.}
\label{fig3}
\end{figure}

In fig.\ref{fig3}a, the entropy $s$ is plotted versus $M$ while keeping $e$ constant: a clear qualitative change  is
detected when going from region I (dashed lines) to region II (solid lines). In the latter domain, the entropy decreases with the 
magnetization $M$, while in the former the entropy grows as $M$ gets larger. In region I, 
hence $(\partial s)/(\partial M)>0$: from relation (\ref{eqh}), and having chosen positive values of $h$, we expect  
$(\partial s)/(\partial e) >0$, which implies working with positive temperatures. 
This prediction is confirmed by visual inspection of fig.\ref{fig3}b: indeed, the entropy $s$ increases with 
the energy $e$, when sampling the portion of the plan associated to region I. More interesting it is instead the scenario that emerges from a close look of region II. Here, $(\partial s)/(\partial M)<0$ and $T^{-1}=(\partial s)/(\partial e) <0$. Summing up the information condensed in figure \ref{fig3}, the temperatures in region I are positive, while those associated to region II are negative. When the critical point $(M_c,e_c)$ is approached the entropy diverges, an observation that suggests targeting region I and II as disconnected domains.  

\begin{figure}
\centering
\includegraphics[width=9cm]{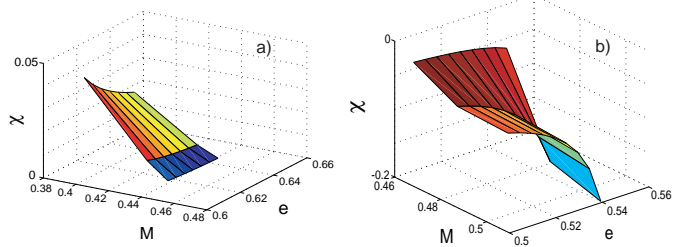}
\caption{The susceptibility $\chi$ is plotted as a function of both $M$ and $e$, in region I (fig. a)) and region II (fig. b)).}
\label{fig4}
\end{figure}

What can we say about the susceptibility $\chi$? The latter can be calculated using relation (\ref{eqchi}). The result of the calculation are reported in figure \ref{fig4}, where the outcome of the analysis in a sub-domain of region I and II is plotted in separate panels. 
As confirmed by direct inspection, in region I the susceptibility is positive, while it is negative in region II. In the latter domain, bound in energy from below, the system thus displays ensemble inequivalence, as revealed by the change in sign of a macroscopically accessible thermodynamic quantity. Intriguingly enough, inequivalence is also accompanied by the additional unconventional condition $T<0$. Coming back to fig.\ref{fig3}, it is worth emphasising that the entropy is a concave function of the energy $e$ (at constant $M$), while it is {\it always} non concave as a function of the magnetization $M$ (at fixed $e$): as we have demonstrated, such a feature is responsible for ensemble inequivalence in region II, but is not sufficient to induce observable inequivalence in region I. 

In this letter we have studied a paradigmatic model of LRI system, perturbed by an external magnetic field. By using a  maximum entropy technique grounded on first principles, we have given both analytical and numerical evidence of (out-of-equilibrium) negative susceptibility. Such a result points to the existence of out-of-equilibrium ensemble inequivalence. According to our analysis, and at difference from previously reported results, the existence of a convex intruder in the entropy does not translate in a sufficient criterion for the negative susceptibility to manifest.

\end{document}